\documentclass[%
 reprint,
 amsmath,amssymb,
 aps,
 prab, 
 floatfix,
]{revtex4-2}
\usepackage{graphicx}
\usepackage{dcolumn}
\usepackage{bm}
\usepackage{hyperref}
\usepackage{xcolor}

\begin{document}

\title{A Living Review Pipeline for AI/ML Applications in Accelerator Physics}

\author{A. Ghribi}
\affiliation{CNRS -- GANIL, Caen, France}
\date{\today}

\begin{abstract}
We present an open-source pipeline for generating a \emph{living review} of artificial intelligence (AI) and machine learning (ML) applications in accelerator physics and technologies. Traditional review articles provide static snapshots that are quickly outdated by the rapid pace of research. The presented system automatically harvests publications from multiple bibliographic sources (arXiv, InspireHEP, HAL, OpenAlex, Crossref, and Springer), deduplicates entries, applies semantic filtering to ensure accelerator and ML relevance, and classifies papers into thematic categories. The resulting curated dataset was exported in JSON, HTML, PDF, and Bib\TeX formats, enabling continuous updates and integration with web frameworks. We describe the methodology, including semantic similarity filtering using sentence-transformer embeddings, threshold calibration, and expert-informed classification. The results demonstrate the robust filtering of $\sim$12000 raw papers/month into a focused corpus of $\sim$2\% relevant works. The pipeline provides the basis for an evolving community-driven review of AI/ML in accelerator science.
\end{abstract}

\maketitle

\section{Introduction}

Artificial intelligence (AI) and machine learning (ML) are reshaping the way scientific research is conceived and conducted. 
In the field of particle accelerators, these methods are increasingly employed for beam dynamics optimisation, feedback control, anomaly detection, surrogate modelling, computer vision--based diagnostics, and reinforcement learning for autonomous tuning. 
The pace of progress is accelerating: while early efforts were isolated demonstrations, entire research programs are now devoted to exploring AI and ML as key enablers of next-generation accelerator design and operation~\cite{Ghribi2025,accml2020survey,reuter2023aiformedaccelerators}.

Traditional review articles, while invaluable for consolidating knowledge, inevitably provide only a \emph{static snapshot} of a rapidly evolving landscape of research. 
Given the exponential growth in AI-related publications and the continuous diversification of methods, such static reviews become outdated within a few months. 
This challenge has already been recognised in high-energy physics (HEP), where the community maintains the \emph{Living Review of Machine Learning for Particle Physics}~\cite{guest2022hepmlreview,radovic2018machinelearninghep}. Ot is worth noting that, alongside the Living Review of Machine Learning for Particle Physics~\cite{hepmllivingreview}, a community fork exists that applies a similar InspireHEP-based workflow to accelerator-related studies\footnote{\url{https://github.com/MALAPA-Collab/AccML-LivingReview}}. Continuously updated resources of this kind, curated by experts, are essential for keeping pace with emerging cross-disciplinary advances. This study introduces a complementary, fully automated, multi-source, and FAIR-compliant pipeline for accelerator science.

Inspired by this model, we present an open-source, fully automated pipeline for generating a living review of artificial intelligence (AI) and machine learning (ML) applications in accelerator physics. 
Unlike traditional reviews that require extensive manual updates, our approach automates the process end-to-end: it harvests publications from multiple bibliographic databases, deduplicates entries, applies semantic filtering to retain only those works at the intersection of accelerators and ML, classifies them into thematic categories, and exports the curated dataset in interoperable formats---JSON, HTML, Bib\TeX, and PDF.

This work aims to provide the accelerator community with the following:
\begin{itemize}
    \item a continuously updated and openly accessible map of AI/ML research relevant to accelerator science;
    \item a transparent and reproducible methodology for literature filtering and classification;
    \item standardized outputs that integrate seamlessly with web platforms, data repositories, and citation workflows.
\end{itemize}

By establishing a living, FAIR-compliant survey of the field, we aim to catalyse collaboration, reduce duplication of effort, and accelerate the responsible adoption of AI/ML methods in accelerator research and operations. The living review is accessible on \texttt{https://aghribi.github.io/acc-ml-living-review/} and contributions remain open through the following repository \texttt{https://github.com/aghribi/acc-ml-living-review}.

\section{System Architecture}

\begin{figure}[t]
    \centering
    \includegraphics[width=\linewidth]{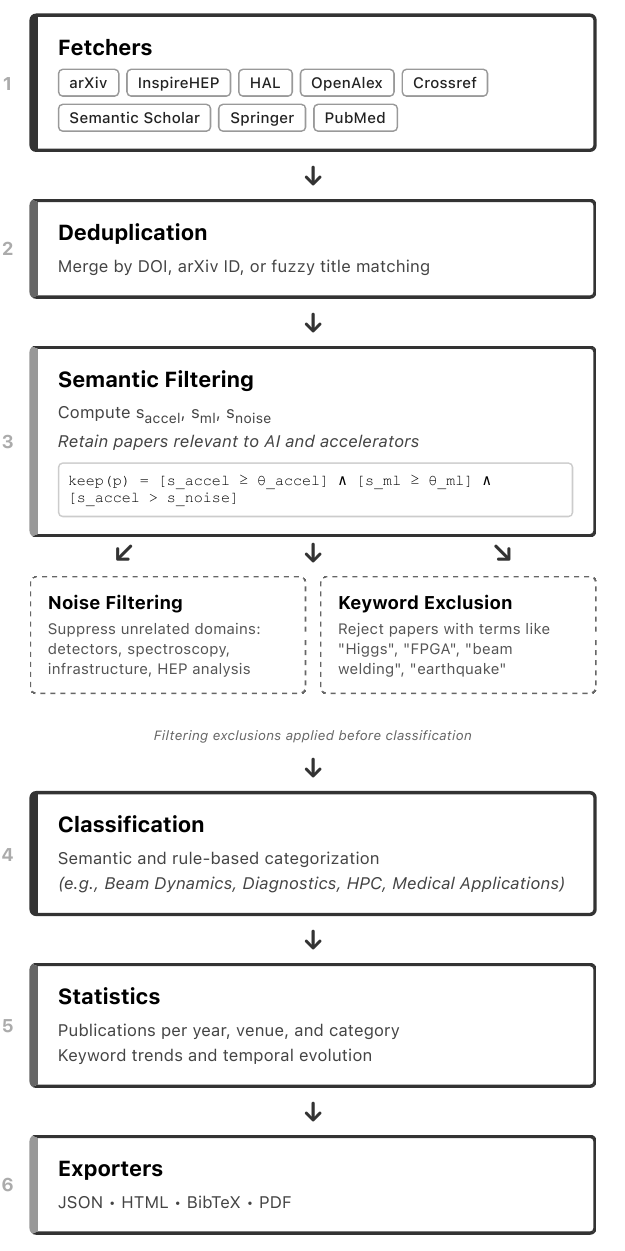}
    \caption{System architecture of the \texttt{living\_review} pipeline.
    Publications are collected from multiple sources, deduplicated, filtered
    by semantic relevance (including noise suppression and keyword
    exclusion), classified, and exported in multiple formats for
    dissemination.}
    \label{fig:pipeline_architecture}
\end{figure}

The \texttt{living\_review} framework is organized as a modular pipeline
(Fig.~\ref{fig:pipeline_architecture}) designed to be robust, extensible,
and reproducible. Each module corresponds to a logically independent task,
allowing future updates or replacement without modifying the rest of the
workflow. The main components are summarised as follows.

\subsection{Fetchers}
Dedicated routines query multiple bibliographic databases relevant to
accelerator physics and engineering, including arXiv, InspireHEP, HAL,
OpenAlex and Crossref. Each fetcher implements a uniform interface and
returns a list of structured \texttt{Paper} objects (title, authors,
abstract, venue, date, identifiers, and links).The This redundancy ensures
broad coverage across physics, engineering, and computer science venues.

\subsection{Deduplication}
Because the same article may appear across several sources (e.g., a preprint
on arXiv, a conference entry in Inspire, and a DOI record in Crossref), a
deduplication step merges duplicates based on canonical identifiers (DOI,
arXiv ID) or fuzzy title matching. The resulting dataset contains only
unique entries.

\subsection{Semantic Filtering and Exclusion}
To retain only works at the intersection of AI/ML and accelerator science,
and to suppress domain noise, the pipeline combines \emph{positive},
\emph{negative}, and \emph{neutral} semantic queries.

Each paper is represented by a joint embedding of its title and abstract
using a lightweight transformer model (\emph{MiniLM})%
~\cite{reimers2020sentencebert}. Three query embeddings are used:
an ``accelerator physics'' reference query, a ``machine learning'' query,
and a ``noise'' query capturing unrelated topics (e.g., particle detectors,
HEP calorimetry, atomic spectroscopy, and infrastructure computing). The
semantic similarity of each paper to these anchors is computed as
$s_{\mathrm{accel}}(p)$, $s_{\mathrm{ml}}(p)$, and $s_{\mathrm{noise}}(p)$.
A paper is retained only if it satisfies the following criteria:
\begin{equation}
\begin{split}
\mathrm{keep}(p) ={}&
    \big[s_{\mathrm{accel}}(p) \ge \theta_{\mathrm{accel}}\big] \\
    &\wedge
    \big[s_{\mathrm{ml}}(p) \ge \theta_{\mathrm{ml}}\big] \\
    &\wedge
    \big[s_{\mathrm{accel}}(p) > s_{\mathrm{noise}}(p)\big].
\end{split}
\end{equation}

This ensures that a paper is simultaneously relevant to both accelerators
and ML, while being semantically distant from the noise domains.

In addition, a curated list of \emph{negative keywords} (e.g., ``Higgs,''
``dark matter,'' ``FPGA accelerator,'' ``beam welding,'' ``earthquake'')
is applied to explicitly remove papers whose content is outside the scope of
accelerator science or pertains to hardware engineering rather than
scientific accelerators. Together, the semantic and keyword filters achieve
high precision in isolating the relevant literature.

\subsection{Classification}
Papers passing the filtering step are classified into thematic categories
such as \emph{Beam Dynamics and Control}, \emph{Diagnostics},
\emph{HPC and Data Management}, \emph{Medical Applications}, and
\emph{Novel Tools and Libraries}. The classifier uses semantic similarity
against predefined category descriptions, complemented by keyword heuristics
and rule-based overrides (e.g. ``surrogate model'' or ``review'' papers).

\subsection{Statistics}
Aggregate statistics provide a global perspective on the field. These include
publication counts per year, per venue, and per category, as well as keyword
trends, and monthly publication dynamics. All metrics are exported alongside
list of papers for downstream visualisation.

\subsection{Exporters}
The curated results were exported in multiple interoperable formats.
\begin{itemize}
    \item \textbf{JSON:} structured data for programmatic access and website integration;
    \item \textbf{HTML:} human-readable, interactive review pages rendered with Jinja2 templating engine;
    \item \textbf{Bib\TeX:} citation-ready files for reference managers;
    \item \textbf{PDF:} automatically generated static summaries using the ReportLab toolkit.
\end{itemize}

This modular design enables continuous updates, straightforward integration
with web frameworks, and reproducible curation of
AI/ML literature on accelerator physics.

\section{Methodology}

The goal of the pipeline is to identify and structure the subset of
scientific literature that lies at the intersection of accelerator physics
and machine learning. Achieving this requires moving beyond keyword
matching toward semantic representations capable of generalizing across
disciplines, and publication venues.

\subsection{Semantic Embeddings}
Each paper $p$ is represented as a dense embedding vector
$\mathbf{e}(p) \in \mathbb{R}^d$ is computed from its title and abstract.
We employ the \texttt{sentence-transformers/all-MiniLM-L6-v2} model
from the \emph{Sentence-Transformers} library%
~\cite{reimers2020sentencebert,minilm2021model}, a lightweight
transformer pretrained on general-purpose natural-language inference and
Semantic textual similarity tasks. The model maps natural-language text
into a semantic space, where related concepts are close in cosine similarity.
Formally,
\begin{equation}
\mathbf{e}(p) =
    f_{\theta}\!\left(\mathrm{title}(p)
    \Vert \mathrm{abstract}(p)\right),
\end{equation}
where $f_{\theta}$ denotes the transformer encoder and
$\Vert$ represents the concatenation.

\subsection{Reference Anchors}
To quantify relevance, the embedding $\mathbf{e}(p)$ is compared to three
reference vectors representing the accelerator, machine-learning, and noise
domains:
\begin{align}
s_{\mathrm{accel}}(p) &=
    \cos\!\left(\mathbf{e}(p), \mathbf{e}_{\mathrm{accel}}\right), \\
s_{\mathrm{ml}}(p) &=
    \cos\!\left(\mathbf{e}(p), \mathbf{e}_{\mathrm{ml}}\right), \\
s_{\mathrm{noise}}(p) &=
    \cos\!\left(\mathbf{e}(p), \mathbf{e}_{\mathrm{noise}}\right).
\end{align}
Here, $\mathbf{e}_{\mathrm{accel}}$ encodes a curated accelerator-physics
reference corpus, $\mathbf{e}_{\mathrm{ml}}$ represents the AI/ML corpus,
and $\mathbf{e}_{\mathrm{noise}}$ captures unrelated domains such as
particle detectors, spectroscopy, high-energy physics analyses, or
infrastructure computing. The cosine similarity is defined as
\begin{equation}
\cos(\mathbf{x}, \mathbf{y}) =
    \frac{\mathbf{x} \cdot \mathbf{y}}
         {\|\mathbf{x}\| \, \|\mathbf{y}\|}.
\end{equation}

\subsection{Thresholding and Exclusion Rule}
A paper is retained only if it is simultaneously relevant to both the
accelerator and ML domains and sufficiently distant from the noise domains:
\begin{equation}
\mathrm{keep}(p) =
    \big[s_{\mathrm{accel}}(p) \ge \theta_{\mathrm{accel}}\big]
    \wedge
    \big[s_{\mathrm{ml}}(p) \ge \theta_{\mathrm{ml}}\big]
    \wedge
    \big[s_{\mathrm{accel}}(p) > s_{\mathrm{noise}}(p)\big].
\end{equation}
Default thresholds
$\theta_{\mathrm{accel}} = 0.13$ and
$\theta_{\mathrm{ml}} = 0.18$
were empirically tuned to balance recall
(capturing genuine AI--for--accelerator papers)
and precision
(excluding tangential or generic ML work).

In addition to the semantic filter, a curated list of
\emph{negative keywords}---including, for example, ``Higgs,''
``calorimeter,'' ``FPGA,'' ``chip,'' ``beam welding,'' and
``earthquake''---is used to explicitly exclude papers outside the scope of
accelerator science, or focused on unrelated hardware engineering. This dual
semantic--lexical filtering strategy minimizes false positives and improves
dataset quality.

\subsection{Classification}
For papers passing the relevance filter, thematic classification is
applied. A hybrid approach was used:
\begin{enumerate}
    \item \textbf{Keyword-assisted rules:} a curated list of accelerator
    and ML terms (e.g., ``beam dynamics'', ``reinforcement learning'',
    ``uncertainty quantification'', ``proton therapy'') is matched against
    titles and abstracts.
    \item \textbf{Semantic clustering (optional):} embeddings may also be
    grouped using unsupervised clustering to reveal emerging topics not yet
    captured using predefined categories.
\end{enumerate}

\subsection{Statistics and Trends}
From the curated corpus, descriptive statistics are computed, including:
\begin{itemize}
    \item publication counts per year and per venue;
    \item category distributions;
    \item keyword frequencies over time;
    \item monthly publication trends.
\end{itemize}
All metrics are exported to structured JSON for downstream visualization on
the Hugo-based website.

\subsection{Export and Reproducibility}
All outputs (JSON, HTML, Bib\TeX{}, and PDF) are automatically generated
from the same curated dataset to ensure reproducibility. The pipeline code
is version-controlled and designed for continuous re-execution as new
publications appear, enabling a fully automated and verifiable living
Review of AI/ML applications in accelerator physics.

The resulting dataset forms the basis for the analyses and case studies presented in Sec. ~\ref{sec:results}.

\section{Results and Case Studies}\label{sec:results}

\subsection{Overall Corpus}
The pipeline was executed on publications spanning 2000–2025, yielding a
curated corpus of $N = 244$ papers at the intersection of machine learning
and accelerator physics. This represents less than $1\%$ of the total
records initially retrieved from arXiv, InspireHEP, HAL, OpenAlex, and
Crossref, underscoring the selectivity and precision of the semantic
filtering process.

\subsection{Temporal Trends}
\begin{figure}[h]
    \centering
    \includegraphics[width=0.48\textwidth]{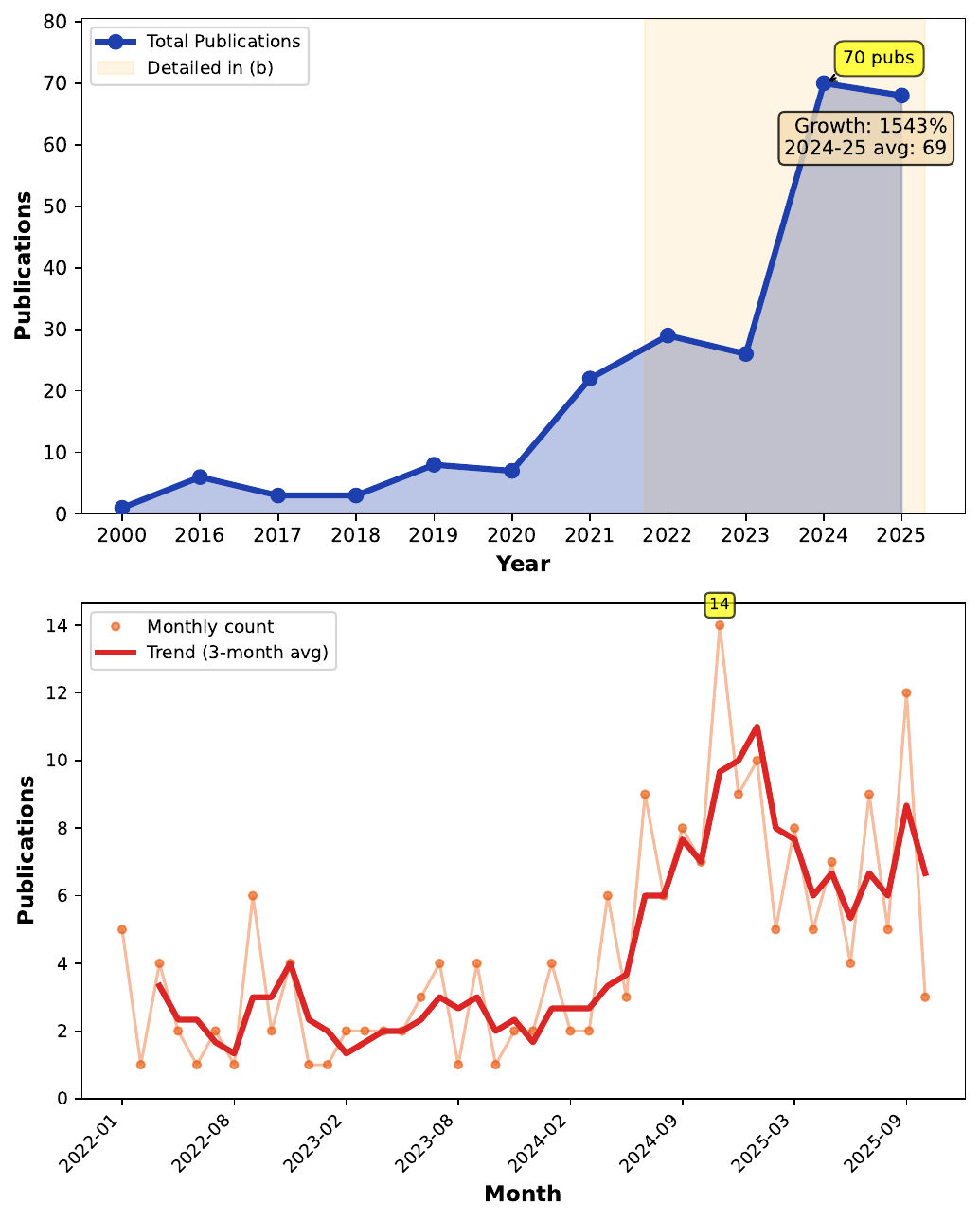}
    \caption{Publication trends analysis showing temporal patterns at two scales: (a) Annual overview from 2000-2025 displaying long-term growth trajectory with significant acceleration from 2021 onwards and peak activity in 2024-2025 (highlighted region); (b) Detailed monthly activity from 2022-2025 revealing short-term fluctuations and sustained productivity with notable peaks, including 3-month moving average trend line for clarity.}
    \label{fig:yearly}
\end{figure}

Figure~\ref{fig:yearly} presents the yearly publication count. The first
identified paper dates to 2000, describing early neural-network-based beam
diagnostics. A pronounced acceleration begins after 2016 and becomes
explosive from 2021 onward, with $70$ and $68$ papers published in 2024 and
2025, respectively. This corresponds to a threefold increase relative to
the 2016--2020 period, confirming the rapid mainstream adoption of AI in
accelerator research.

\subsection{Category Distribution}
\begin{figure}[h]
    \centering
    \includegraphics[width=0.48\textwidth]{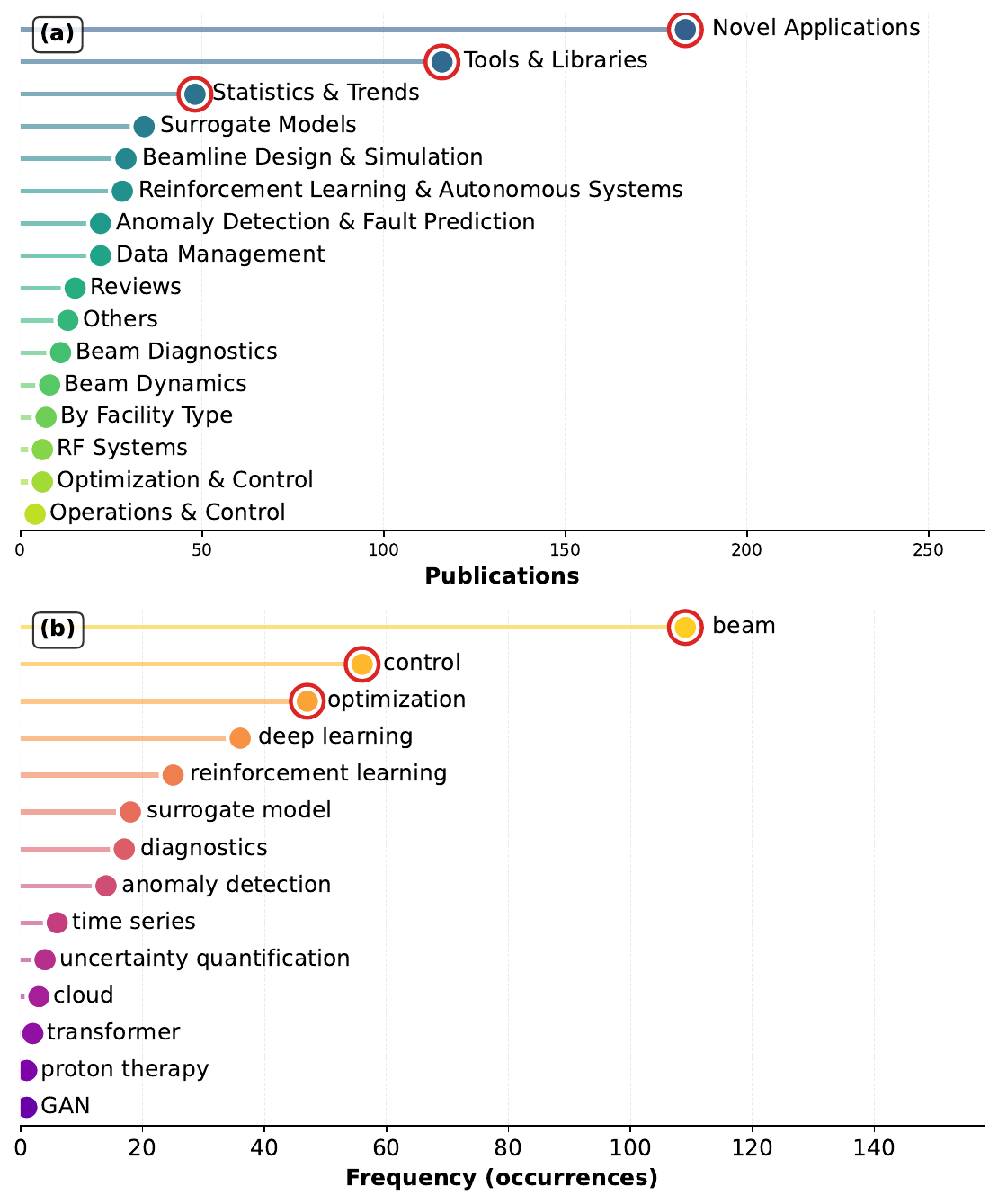}
    \caption{Thematic distribution: (a) Research categories ranked by publication count ;
(b) keyword frequency distribution revealing specific research topics.}
    \label{fig:categories}
\end{figure}

Automatic classification revealed 16 thematic groups. The most
populated are:
\begin{itemize}
    \item \textbf{Novel Applications} (183 papers, $\sim35\%$) --- cross-domain
    uses of AI within accelerator environments, including simulation
    surrogates, anomaly detection, and experimental optimisation.
    \item \textbf{Tools and Libraries} (116) --- reusable software
    frameworks, datasets, and benchmark platforms.
    \item \textbf{Surrogate Models} (34) --- fast emulators of beamline and
    RF components;
    \item \textbf{Reinforcement Learning and Autonomous Systems} (28) ---
    adaptive control and tuning strategies
    \item \textbf{Beamline Design and Simulation} (29) --- optimization of
    optical lattices and injection systems.
\end{itemize}
Smaller but active areas include \emph{Anomaly Detection and Fault
Prediction} (22), \emph{Data Management} (22), and \emph{Reviews} (15).
Figure~\ref{fig:categories} shows the relative proportions of

\subsection{Keyword Analysis}
Keyword frequency analysis confirms the prominence of ``beam'' (109
occurrences), ``control'' (56), and ``optimization'' (47) as central
themes, reflecting the field’s strong focus on data-driven tuning and
beamline performance. Machine-learning methods such as ``deep learning''
(36) and ``reinforcement learning'' (25) appear increasingly in control and
diagnostics studies. Emerging topics include ``transformers'',
``foundation models'', and ``federated learning'', indicating the diffusion
of advanced AI architectures in accelerator contexts. Mentions of
``proton therapy'' and ``GAN'' highlight the growing overlap with medical
and imaging application.

\subsection{Recent Momentum}
Monthly publication dynamics reveal sustained high activity throughout
2024--2025, with peaks in November~2024 (14 papers) and
September2025 (12). This consistency suggests the formation of a
maturing and expanding research community at the AI--accelerator
interface, supported by cross-institutional collaborations and open
data-sharing efforts.

\subsection{Case Studies}
Three representative case studies illustrate the diversity of approaches.
\begin{enumerate}
    \item \textbf{Beam tuning with reinforcement learning:} actor--critic
    and policy-gradient algorithms have been applied to storage-ring and
    linac tuning, achieving order-of-magnitude reductions in setup time
    compared with the manual procedures.
    \item \textbf{Medical quality assurance with generative models:}
    adversarial and diffusion networks perform MRI-to-CT translation for
    proton-therapy dose prediction and reducing patient imaging requirements.
    \item \textbf{Surrogate modeling of RF and beamline components:} deep
    neural networks trained on high-fidelity electromagnetic simulations
    yield millisecond-scale predictions of cavity fields and beam optics,
    This enables interactive design exploration.
\end{enumerate}

\subsection{Comparison to HEP-ML Living Review}
Unlike the HEP-ML Living Review~\cite{hepmllivingreview}, which provides a
manually curated overview of machine-learning use across high-energy
physics, the present pipeline targets the accelerator domain specifically
and automates the entire curation processes. This enables near-real-time
updates, quantitative trend analysis, and seamless integration with FAIR
data infrastructures.

\section{Discussion}

\subsection{Strengths of the Pipeline}
The \texttt{living\_review} pipeline provides a scalable and reproducible
as an alternative to traditional manually curated surveys. Its modular design
and integration of multiple bibliographic APIs (arXiv, InspireHEP, HAL,
OpenAlex, Crossref) enable the automatic retrieval and filtering of
thousands of publications in minutes. By relying on semantic
embeddings rather than keyword matching, the system robustly identifies
works situated at the accelerator--ML interface, reducing bias and
capturing cross-disciplinary research that conventional search strategies
might overlook. Because the entire process is automated, the review can be
regenerated on a regular basis, maintaining a continuously updated and
FAIR-compliant record of research activity that integrates seamlessly with
Open science infrastructure.

\subsection{Limitations}
Several limitations highlight the need for improvement. The use of
fixed semantic thresholds
($\theta_{\mathrm{accel}} = 0.13$ and $\theta_{\mathrm{ml}} = 0.18$)
introduces sensitivity to boundary cases, occasionally excluding relevant
papers or admitting tangential ones. Some degree of source bias persists,
as InspireHEP and Crossref remain dominant, while arXiv preprints are
frequently merged or overwritten during deduplication, leading to partial
underrepresentation of the preprints. In addition, incomplete metadata—such as
missing abstracts, inconsistent author identifiers, or absent DOIs—can
This limits the accuracy of trend analysis and classification.

\subsection{Comparison with Manual Reviews}
Relative to human-curated reviews, the automated approach optimizes for
\emph{completeness}, \emph{timeliness}, and \emph{reproducibility}, albeit
with a reduced contextual precision. Expert-led reviews remain invaluable
for detailed methodological interpretation, whereas automated curation
It offers a scalable and unbiased quantitative backbone. In practice, both
approaches are complementary: automation ensures broad and reproducible
coverage, whereas expert judgment provides depth and validation.

\subsection{Future Directions}
Several enhancements are planned to extend both the robustness and
interpretability of the framework
\begin{itemize}
    \item \textbf{Model improvements:} Integration of larger
    transformer-based language models with domain-adapted embeddings to
    enhances semantic discrimination and reduces misclassification.

    \item \textbf{Calibration and benchmarking:} Validation of
    classification thresholds using expert-annotated subsets and creation
    of benchmark datasets to measure reproducibility and precision across
    releases.

    \item \textbf{Hierarchical taxonomy:} Development of a multi-level
    category structure (e.g., \emph{Beam Physics} $\rightarrow$
    \emph{Control}, \emph{Diagnostics}, \emph{Simulation}) to reflect
    thematic hierarchies and subfields.

    \item \textbf{Explainability:} Implementation of SHAP-based and
    centroid-based visualizations to interpret classification outcomes and
    provide transparent insights into category assignments.

    \item \textbf{Uncertainty-aware analytics:} Use of probability-weighted
    statistics to visualize uncertainty and confidence in category trends
    within the Living Reviews dashboard.

    \item \textbf{Extended coverage:} Inclusion of additional bibliographic
    and grey-literature sources (e.g., technical design reports, conference
    proceedings, and internal notes) to broaden the corpus completeness.
\end{itemize}

\section{Conclusion and Outlook}

We have introduced the \textbf{Living Review pipeline}, an open and
reproducible framework for the automated collection, filtering, and
classification of research at the intersection of accelerator physics and
machine learning. The system integrates multiple bibliographic sources,
semantic filtering, and lightweight categorization to produce a
machine-curated, FAIR-compliant survey of the field that can be continuously
updated as new publications are released.

This work complements the community-driven
\emph{Living Review of Machine Learning for Particle Physics}%
~\cite{hepmllivingreview} and its accelerator-focused derivatives, which
rely on InspireHEP-based workflow. The present framework extends these
efforts by introducing a fully automated, multi-source, and
semantic-driven approach designed specifically for the accelerator science.

As accelerators become increasingly complex and data-rich, and as AI
techniques become integral to their operation, diagnostics, and design,
such tools are vital to maintaining a comprehensive and transparent view of
the evolving research landscape. Beyond cataloguing progress, the
\texttt{living\_review} framework provides a quantitative foundation for
evidence-based policy, collaborative research planning, and the
responsibly integrate AI methods across the accelerator ecosystem.

\begin{acknowledgments}
The author thanks the developers of the HEP-ML Living Review for inspiring this work and acknowledges the open-source tools and APIs (InspireHEP, arXiv, HAL, OpenAlex, and Crossref) that made this project possible.

\end{acknowledgments}

\bibliographystyle{apsrev4-2}
\bibliography{ref}

\end{document}